# Mitigating Backdoor Attack Via Prerequisite Transformation

Han Gao

Changchun University of Technology

## Abstract:

In recent years, with the successful application of DNN in fields such as NLP and CV, its security has also received widespread attention. (Author) proposed the method of backdoor attack in Badnet. Switch implanted backdoor into the model by poisoning the training samples. The model with backdoor did not exhibit any abnormalities on the normal validation sample set, but in the input with trigger, they were mistakenly classified as the attacker's designated category or randomly classified as a different category from the ground truth, This attack method seriously threatens the normal application of DNN in real life, such as autonomous driving, object detection, etc.This article proposes a new method to combat backdoor attacks. We refer to the features in the area covered by the trigger as trigger features, and the remaining areas as normal features. By introducing prerequisite calculation conditions during the training process, these conditions have little impact on normal features and trigger features, and can complete the training of a standard backdoor model. The model trained under these prerequisite calculation conditions can, In the verification set $D'_{val}$ with the same premise calculation conditions, the performance is consistent with that of the ordinary backdoor model. However, in the verification set $D_{val}$ without the premise calculation conditions, the verification accuracy decreases very little (7%~12%), while the attack success rate (ASR) decreases from 90% to about 8%.Author call this method Prerequisite Transformation(PT).

## 1 Introduction:

In real life, users often use models or data provided by third parties to achieve their goals when training models due to insufficient computing power or insufficient data sets. This provides an opportunity for attackers to attack by changing the training data to train or provide a model with a backdoor, To reduce the accuracy of the model or force it to misclassify with or without targets.

Backdoor attacks were first proposed in a paper by XX in XX year. By adding triggers to the training data, the attacked model is trained on this data set denoted by $D_{poisoned}$ to generate a model $F_{trigger}$ ($F_t$) with backdoor. This backdoor model will perform well on a normal data set, , it will misclassified to the category specified by the attacker or to a different category than Ground Truth,while on inputs with backdoor triggers. The representation of a Trigger is a certain patch in the image, such as the white patch in the bottom right corner of Fig1.

The backdoor attack has a significant property, which is also the essence of the neural network's strong learning ability about the feature of the trigger location, that is, it is caused by over fitting. We call the feature of the trigger location as trigger feature, while the area not covered by the trigger is called normal feature. This paper proposes an idea based on this - to add some prerequisites during the model learning process, which have a small impact on the model's learning of normal features, while the triggering features have a small impact on the training process due to their backdoor attack characteristics. Training the model on this condition to obtain the backdoor model Ft. In subsequent experiments, it can be observed that adding the same prerequisite as training data set to the validation data set does not decrease the success rate (ASR) of backdoor attacks. However, when removing the prerequisite on the validation data set, it can be observed that ASR decreases sharply (from 90% to 8%), while the performance of the model in normal samples does not decrease significantly (7% to 12%).

The viewpoint proposed in Rethinking of Trigger[1] is that when the position of the trigger changes or the value of the trigger decreases, ASR will be reduced. Based on this viewpoint, this article proposes a processing method on training data (regardless of whether it is poisoned or not), which greatly reduces the impact of backdoors and the success rate of backdoor attacks by processing the training data without a decrease in accuracy.

# 2 Introduction to backdoor attacks:

Backdoor attack mainly refers to the operation of the attacker implanting a trigger into the training samples during the model training process, in order to achieve the same performance of the model on samples without trigger compare to the accuracy of the model obtained from ordinary training without implanting a trigger into the training sample. For sample inputs with triggers, the backdoor model outputs random or fixed categories that adversary desired.

Backdoor attacks can be specifically divided into the following types of attack methods based on the way they are poisoned,

## 2.1 Layout of triggers:

The layout of triggers can be divided into local triggers and global triggers. Local triggers can refer to the method in "Badnet[2]", which covers a small area of the sample with fixed area and eigenvalues.The triggered sample is shown in the Fig.1. The global trigger is a method proposed in "Blend"[3] that takes the entire sample as the target and inserts a fixed trigger into the sample through interpolation. The sample (X, Y) is denoted as inputs and its label, and the sample X with the trigger is $X_t$, and the trigger is T, hyper-parameters $\alpha \in (0,1)$, then the implantation method of the global trigger is $X_t=(1-\alpha) X+ \alpha T$. The triggered sample is shown in the Fig.2.

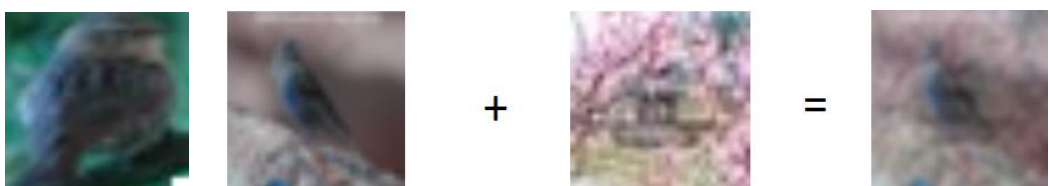

Fig.1  
Local trigger

Fig.2  
Globe trigger

## 2.2 the way to poison:

From the perspective of selecting poisoned samples, it can be further divided into two types: "clean label"[4] attack and "non clean label" attack. Clean label attack aim to selecting samples for poisoning, whose ground truth is consistent with the target category. we denote gt(x) is the ground truth of the sample x, and the sample (X, Y) is denoted as inputs and its label,and the sample X with a trigger is $X_t$. The clean sample attack meet the requirement gt ($X_t$)=target label. Non clean label attacks do not have the limitations of clean sample attacks, and the ground truth of $X_t$ under this attack method can be the same as or different from the target label.

## 3 Related work:

In [5] proposed the utilization of data augmentation (spatial transformation) to reduce the impact of backdoor attacks by changing the position and value of triggers.[6] proposed a approach that pruning the neural network to eliminate neurons affected by triggers and reduce the impact of backdoor attacks on the model. This method requires a model with a backdoor to be trained and pruned on a validation data set $D_{val}$ where the attacker does not have permission.In [7]strong data augmentation, data augmentation methods such as Mixup and CutMix are used to combat backdoor attacks.Contributions of this paper are followed:

(1) This article proposes a new approach to combat backdoor attacks using local triggers.(2) witch has low computational complexity and can be used in real-time tasks.(3) and does not require retraining, fine-tuning, or pruning on clean samples.(4) Deployment of this approach is quiet simple.

## 4 Experimentation:

### 4.1 Data-set:

We use data set Cifar-10 witch has 10 categories, each containing a total of 5000 samples ,totaling 50000 samples. We conducted two poisoning methods. (1)Using a non label clean approach, 250 samples were randomly selected from each category, with a total of 2500 samples accounting for 5% of the whole data set, and placed in the target category, called bird in this paper. (2)Using a label clean method, randomly select 2500 samples from the target label, accounting for 5% of the total data set for poisoning.

### 4.2 Method:

The author of this paper conducted experiments using Mixup and CutMix to mitigate the effect of backdoor attack and found that in the case of non clean labels attack, the accuracy (ACC) on benign sample and attack success rate (ASR) were both high in the second half of the training. From result we can see the 5 to 30 Epoch both ACC and ASR maintained a relatively high accuracy, while the training results of high ACC and low ASR occurred in the previous rounds of training, in the 0 to 4 Epoch, as shown in Fig3.1 & Fig3.2. This result also confirms that the learning of the model on trigger is surprising, that is, over fitting. Under the clean label attack method, both methods performed very stably, as shown in Fig4.1 & Fig4.2.

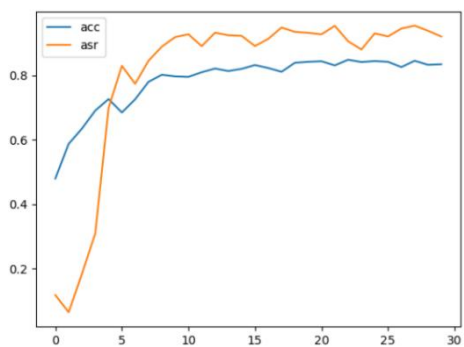
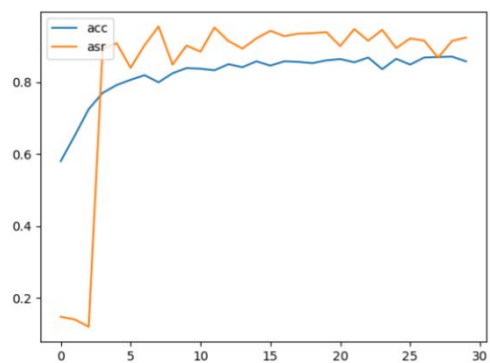

| Fig3.1 | Fig3.2 |
| --- | --- |
| Mixup train on data of non-clean-label attack | CutMix train on data of non-clean-label attack |

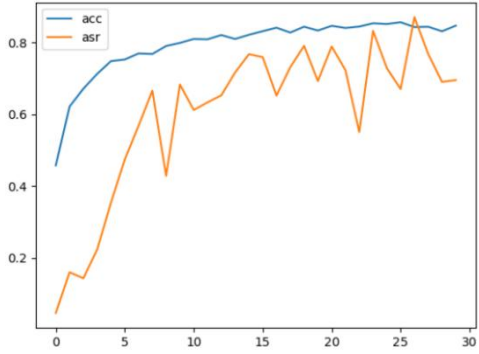 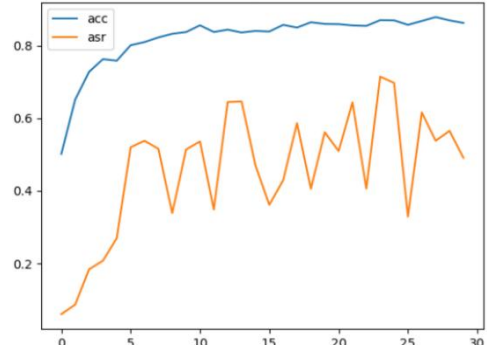

| Fig4.1 | Fig4.2 |
| --- | --- |
| Mixup train on data of clean-label attack | CutMix train on data of clean-label attack |

The performance of [6,7] and the method proposed in this article in combating backdoor attacks is shown in Table 1. It can be seen that the method proposed in this article requires less time when it has a similar defense effect against backdoor attacks, which means that the method proposed in this article has some advantages in real-time critical tasks.

[6] assume that the user holds a data set $D_{val}$ with no permissions for adversarial, and prunes the model during the process of retraining the model with $D_{val}$. If there is no $D_{val}$ assumed in the text, this method will fail.But we assume that we adopt data set or module from third party, advisory has the most access to the data set of validation and training, our approach still works well in this situation.But we assume that we adopt data-set or module from third party,adversarial have the most access to the data-set of validation and training,our approach still works well in this situation.

The method adopted in this paper is to randomly collect and average training data (triggered or not) to obtain a position near the center point in the feature space of the training data-set. By adding and subtracting, the overall feature space of the training data is pulled in or out, corresponding to reducing or increasing the feature gap between samples in each category. By using mod to compress the feature values of the input samples in a certain region, the goal of reducing the trigger value is achieved. As a prerequisite transformation, we set this condition as Reflector (R), and let the input X pass through R to obtain X ', that is, R (X)=X'.Refractor is defined as below：

$$(X-(average(\sum X_{select})-X)/Ratio) \bmod P \qquad (1)$$

$X_{select}$ is denoted as that X sampled number of n from all class in training set,P and Ration is the empirical setting.

Then, X 'is input as a new sample into model F to train the model $F_r$. At this time, the accuracy of the training process is shown in Fig.5, and the performance of $F_r$ is shown in Table1 and 2. Through the observation of Fig.5, we can see that the ASR in the first epoch is very low, but it will increase sharply to a large value in the next epoch, indicating that the model has fully learned the trigger in this epoch. At the same time, it is observed that the changes in ACC are not significantly different from the ACC during normal training, which means that the added prerequisite transformation have little impact on the learning of normal features. During the testing phase, the input $X_{test}$ of the model will not undergo R changes, and the performance of the model $F_r$ in this situation is shown in TABLE1 and TABLE2 in test_acc and test_asr column. Through TABLE1 and 2, it can be observed that the ACC of the model is almost unaffected without any prerequisite transformation, while the ASR decreases sharply. By observing this phenomenon, we can change the training conditions by adding a Refractor proposed in this paper when using untrusted third-party data for training. The model generated on this basis will not modify the input through Refractor when put into use to achieve normal utilization. The idea of this paper comes from the phenomenon that the models are easy to produce overfitting to the trigger in the training, and author uses this phenomenon reasonably to ensure the correct deployment of the model in life to a certain extent, and reduces the harm of the back door in the use of the model through a plain way. To my best knowledge, I am the first person to adopt this method to mitigate backdoor attack.

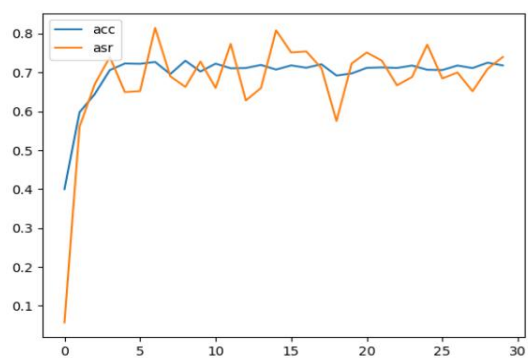

Fig.5

| clean_label, local trigger | | avg_acc | avg_asr | best_acc | asr of best_acc | test_acc | test_asr | time |
|---|---|---|---|---|---|---|---|---|
| Baseline | resnet | 0.7754 | 0.80915 | 0.8055 | 0.8158 | 0.8065 | 0.8158 | 1198 |
| Refractor(Ours) | resnet | 0.69686 | 0.67892 | 0.7303 | 0.6629 | 0.7246 | 0.0706 | 1303 |
| Mixup | resnet | 0.793153 | 0.602236 | 0.8564 | 0.6701 | 0.8678 | 0.6701 | 1680 |
| CutMix | resnet | 0.8213 | 0.45793 | 0.8789 | 0.538 | 0.8746 | 0.4159 | 1592 |

TABLE1

test_acc and test_asr in Baseline, Mixup and CutMix are come from normally, witch means the testing data do not undergo R changes, besides testing data of Refractor(Ours)

| non_clean_label, local trigger | | avg_acc | avg_asr | best_acc | asr of best_acc | test_acc | test_asr | time |
|---|---|---|---|---|---|---|---|---|
| Baseline | resnet | 0.78442 | 0.95116 | 0.8037 | 0.9571 | 0.8037 | 0.9571 | 1193 |
| Rafractor(Ours) | resnet | 0.696063333 | 0.68682 | 0.732 | 0.6342 | 0.6662 | 0.0886 | 1278 |
| mixup | resnet | 0.783813333 | 0.80454 | 0.8488 | 0.9062 | 0.8639 | 0.9062 | 1782 |
| cutmix | resnet | 0.82298 | 0.8345 | 0.8704 | 0.914 | 0.8627 | 0.8389 | 1426 |

TABLE2

test_acc and test_asr in Baseline, Mixup and CutMix are come from normally, witch means the testing data do not undergo R changes, besides testing data of Refractor(Ours)

## 5 Conclusion:

Based on the characteristics of backdoor attacks, that is, the model's overfitting learning of triggers, this paper proposes a calculation-saving method to mitigate the impact of backdoor attacks. Moreover, the assumptions in this paper are very simple, and do not require additional completely clean data sets to retrain,fine tune or pruning, which makes the scenarios used in this method more realistic. Fewer calculations also make it more suitable for "real time critical" tasks.

We will focus our attention on the trigger feature ($T_f$) where the trigger is located. From formula （1）, we can conclude that

$$T_f' = (T_f - (\text{average}(\sum X_{select}) - T_f)/\text{Ratio}) \bmod P \tag{2}$$

$$T_f' = (1+\text{ratio}) * T_f \text{ average}(\sum X_{select})/\text{ration}) \bmod P \tag{3}$$

Since the mapping point in this article is fixed, set it to A, which is

$$T_f' = (1+\text{ratio}) * \text{trigger A})/\text{ratio}) \bmod P \tag{4}$$

It is easy to think through observation that if a value of $T_f = T_f'$ is found, this method will fail. In order to better address this issue, when selecting a Refractor, based on formula (1), the selected points can be closer to a certain class or classes. This method looks more like a random number operation in information transmission security. Through this idea, we may be able to extend the application of symmetric or asymmetric encryption in deep learning security.

## 6 Future work:

Through experiments, the method proposed in this article is highly effective in defending against backdoor attacks using local triggers (whether they are clean label attacks or non clean label attacks), but has almost no effect on global trigger backdoor attacks (Blend). The current value of 'Refractor' is fixed, manifested as a point at the center of the feature space. In the future, attempts will be made to bring the mapping points passing through the Refractor closer to a class from the training data set to reduce attacks against the method proposed in this paper.Although the method proposed in this article performs well in backdoor attacks on local triggers,but it performs poorly in defense against backdoor attacks on global triggers. (1) Future work will focus on how to mitigate the impact of this attack through the approach presented in this article. (2) Identify and apply methods from past information security that can be applied to deep learning security, such as symmetric encryption and asymmetric encryption.

## REFERENCE:

[1] Li, Y., Zhai, T., Wu, B., Jiang, Y., Li, Z., & Xia, S. (2020). Rethinking the trigger of backdoor attack. *arXiv preprint arXiv:2004.04692*.

[2 ]Gu, T., Dolan-Gavitt, B., & Garg, S. (2017). Badnets: Identifying vulnerabilities in the machine learning model supply chain. *arXiv preprint arXiv:1708.06733*.

[3] Chen, X., Liu, C., Li, B., Lu, K., & Song, D. (2017). Targeted backdoor attacks on deep learning systems using data poisoning. *arXiv preprint arXiv:1712.05526*.

[4] Turner, A., Tsipras, D., & Madry, A. (2019). Label-consistent backdoor attacks. *arXiv preprint arXiv:1912.02771*.

[5] Qiu, H., Zeng, Y., Guo, S., Zhang, T., Qiu, M., & Thuraisingham, B. (2021, May). Deepsweep: An evaluation framework for mitigating DNN backdoor attacks using data augmentation. In *Proceedings of the 2021 ACM Asia Conference on Computer and Communications Security* (pp. 363-377).

[6] Liu, K., Dolan-Gavitt, B., & Garg, S. (2018). Fine-pruning: Defending against backdooring

attacks on deep neural networks. In *Research in Attacks, Intrusions, and Defenses: 21st International Symposium, RAID 2018, Heraklion, Crete, Greece, September 10-12, 2018, Proceedings 21* (pp. 273-294). Springer International Publishing.

[7] Borgnia, E., Cherepanova, V., Fowl, L., Ghiasi, A., Geiping, J., Goldblum, M., ... & Gupta, A. (2021, June). Strong data augmentation sanitizes poisoning and backdoor attacks without an accuracy tradeoff. In *ICASSP 2021-2021 IEEE International Conference on Acoustics, Speech and Signal Processing (ICASSP)* (pp. 3855-3859). IEEE.